%% file: gradient-revise3.tex
\documentclass[prl, superscriptaddress,twocolumn]{revtex4} %,preprint,a4paper
\usepackage{amsmath,amsthm,amsfonts,amssymb,bm}
\usepackage{graphicx}

\begin{document}

\title{Storage of multiple coherent microwave excitations in an electron spin ensemble}

\author{Hua Wu}
\affiliation{Department of Materials, Oxford University, Oxford OX1 3PH, UK}

\author{Richard E. George}
\affiliation{Clarendon Laboratory, Department of Physics, Oxford University, Oxford OX1 3PU, UK}

\author{Janus H. Wesenberg}
\affiliation{Department of Materials, Oxford University, Oxford OX1 3PH, UK}

\author{Klaus M\o lmer}
\affiliation{Department of Physics and Astronomy, University of Aarhus, DK-8000 Aarhus
C., Denmark}

\author{David I. Schuster}
\affiliation{Departments of Applied Physics and Physics, Yale University, New Haven, Connecticut 06520, USA}

\author{Robert J. Schoelkopf}
\affiliation{Departments of Applied Physics and Physics, Yale University, New Haven, Connecticut 06520, USA}

\author{Kohei M. Itoh}
\affiliation{School of Fundamental Science and Technology, Keio University, Yokohama 223-8522, Japan}

\author{Arzhang Ardavan}
\affiliation{Clarendon Laboratory, Department of Physics, Oxford University, Oxford OX1 3PU, UK}

\author{John J. L. Morton} 
\affiliation{Department of Materials, Oxford University, Oxford OX1 3PH, UK}
\affiliation{Clarendon Laboratory, Department of Physics, Oxford University, Oxford OX1 3PU, UK}

\author{G. Andrew D. Briggs}
\affiliation{Department of Materials, Oxford University, Oxford OX1 3PH, UK}

\date{\today}

\begin{abstract}
Electron and nuclear spins have good coherence times and an ensemble of spins is a promising candidate for a quantum memory. By employing holographic techniques via field gradients a single ensemble may be used to store many bits of information. Here we present a coherent memory using a pulsed magnetic field gradient, and demonstrate the storage and retrieval of up to 100 weak 10~GHz coherent excitations in collective states of an electron spin ensemble. We further show that such collective excitations in the electron spin can then be stored in nuclear spin states, which offer coherence times in excess of seconds.
%AStrong coupling between a microwave photon and electron spins, which could enable a long-lived quantum memory element for superconducting qubits, is only possible if a large ensemble of spins is used. This represents an inefficient use of resources unless multiple photons, or qubits, can be orthogonally stored and retrieved, for example using holographic techniques.  recent paper [J. H. Wesenberg \emph{et al.}, Phys. Rev. Lett. \textbf{103}, 070502 (2009)] has proposed a new quantum computing scheme using electron spin ensemble as the storage medium. In this scheme multiple quantum bits are encoded in different collective excitations of the electron spins. Here we present the first experiment on the storage and retrieval of multiple weak 10~GHz coherent excitations in distributed memories based on electron spin ensembles. We further demonstrate the use of a coupled nuclear spin, which can offer coherence times in excess of seconds, for even more robust storage.
\end{abstract}

\maketitle

Instead of storing information in specific locations as in photography and in conventional computer memory, information can be stored in distributed collective modes, as in holography. Advantages include obviating the need for local manipulations and measurements, enhanced coupling to electromagnetic fields, and robustness against decoherence of individual members of the ensemble. This principle has been applied to different light-matter interfaces such as atoms~\cite{Nunn2008, Hetet2008, Vasilyev2008, Hosseini2009}, ion-doped crystals~\cite{Hetet2008a, Nilsson2005, Alexander2006, Kraus2006, Riedmatten2008}, polar molecules \cite{Rabl2006, Tordrup2008, Andre2006, Tordrup2008a}, or spins~\cite{Wesenberg2009, Imamoglu2009}. Controlled reversible inhomogeneous broadening (CRIB)~\cite{Kraus2006}, or gradient echo memory (GEM)~\cite{Alexander2006} schemes which apply external field gradients to address different storage modes have been proposed and observed in gaseous atomic samples~\cite{Nunn2008,Hetet2008} and in ion-doped solids~\cite{Nilsson2005,Alexander2006}. 

In this Letter, we demonstrate the storage of multiple microwave excitations in an electron spin ensemble. The spin ensembles used as the storage medium are the electron spin of nitrogen atoms in fullerene cages ($^{14}$N$@$C$_{60}$) and phosphorous donors in silicon. The microwave excitations are phase encoded using a static or pulsed field gradient, with the latter allowing for recall in arbitrary order. We have stored up to 100 weak excitations in a spin ensemble and recalled them sequentially. We also demonstrate the coherent transfer of the stored multiple excitations between electron spin and nuclear spin, which will allow much longer storage times~\cite{Morton2008}. The multi-mode storage  achieved in this way offers prospects of constructing a long-lived quantum memory which could be used for a hybrid quantum computing architecture with superconducting qubits.

%\section{Magnetic field gradient and qubit encoding}

A quasistatic magnetic field along the $z$-axis causes the members of the spin ensemble to precess at an angular frequency $B(z,t)\mu g_e/\hbar$, where $\mu$ is the Bohr magneton and $g_e$ is the electron gyromagnetic ratio. Applying a magnetic gradient of strength $G=\partial{B(z,t)}/{\partial z}$ for a time $\tau$ consequently leads to a difference in precession angle of $\delta \theta=(\mu g_e/\hbar)G\tau\cdot\delta z$ between two spins with separation $\delta z$ along the $z$-axis. A gradient pulse thus maps a spin state with a coherent transverse magnetization (such as that generated by a global resonant microwave tipping pulse) to a spin-wave excitation associated with a wave number $k=(\mu g_e/ \hbar)G\tau\cdot z$. Each further application of $G$ for duration $\tau$ generates a change in the wavevector of the global spin wave mode, by an amount $k$. The notion of one or two dimensional $k$-space introduced by magnetic field gradients has been widely used for a variety of nuclear magnetic resonance (NMR) applications~\cite{Hahn1955,Sodickson1998} including imaging~\cite{Gossuin2010}. 

Using this approach it is possible to store many coherences in the electron spin ensemble. If a global coherence is generated by a small electron spin resonance (ESR) tipping pulse $\delta\theta\ll\pi$, the amplitude of the coherence (given by the magnetization in the $x$-$y$ plane) is proportional to $\sin(\delta\theta) \sim \delta\theta$, while the change in the magnetization along $z$ is given by $1-\cos(\delta\theta) \sim \delta\theta^2$ which can be neglected. A gradient pulse converts this $k_i=0$ coherence into a state $k_f\neq0$. These two modes are orthogonal if $\frac{1}{N} \int_{-d/2}^{d/2} n(z)e^{-i(k_i-k_j)\cdot z}\textrm{d}z=0$, where $n(z)$ is the density of spins at position $z$, $N$ is the total number of spins, $d$ is the extent of the sample along the $z$ direction. This condition is satisfied if the total magnetization in the $x$-$y$ plane is zero for the $k_f$-mode; this ensures that the $k_f$-mode does not radiate as the ensemble precesses in the static magnetic field. Since, to first order, the total $z$-axis magnetization is unchanged by the small tipping pulse, a new $k=0$ coherence may be generated by a further small tipping pulse, and stored in another $k\neq0$ mode by a further gradient pulse.

Information can be encoded on these modes by choosing the phase of the small tipping pulses, i.e.~the axis of the tip in the $x$-$y$ plane. To retrieve the information stored in mode $k$, one needs a change in wave number by the amount $-k$ to bring it back to $k=0$ mode. This is accomplished by applying the opposite field gradient $-G$ for the same duration $\tau$. The net magnetisation of this mode precesses in the external field causing it to radiate a signal similar to a spin echo~\cite{Schweiger&Jeschke}. In our experimental setup, we generate the pulsed magnetic field gradient along the $z$-axis by placing an anti-Helmholtz coil outside the resonator. The current pulses used to produce the field gradient pulse are highly reproducible. In addition to the controlled inhomogeneity of the gradient pulses, there is a random inhomogeneity caused by variations in the environment of the individual spins of the ensemble and some residual gradients in the static applied magnetic field. The effect of this uncontrolled but static inhomogeneity can be refocussed using a Hahn echo sequence~\cite{Schweiger&Jeschke}. A $\pi$-pulse echo also has the effect of inverting the wave vectors of stored collective excitations $k\to -k$. Thus, following an odd number of refocusing $\pi$-pulses, a field gradient of the original polarity $G$ restores a $k$-mode to the $k = 0$ mode.

%\subsection{Experimental setup}

%\subsection{Quantum phase coherence eliminated and recovered by gradient pulse(s)}

\begin{figure}[t]
\includegraphics[width=3.5in]{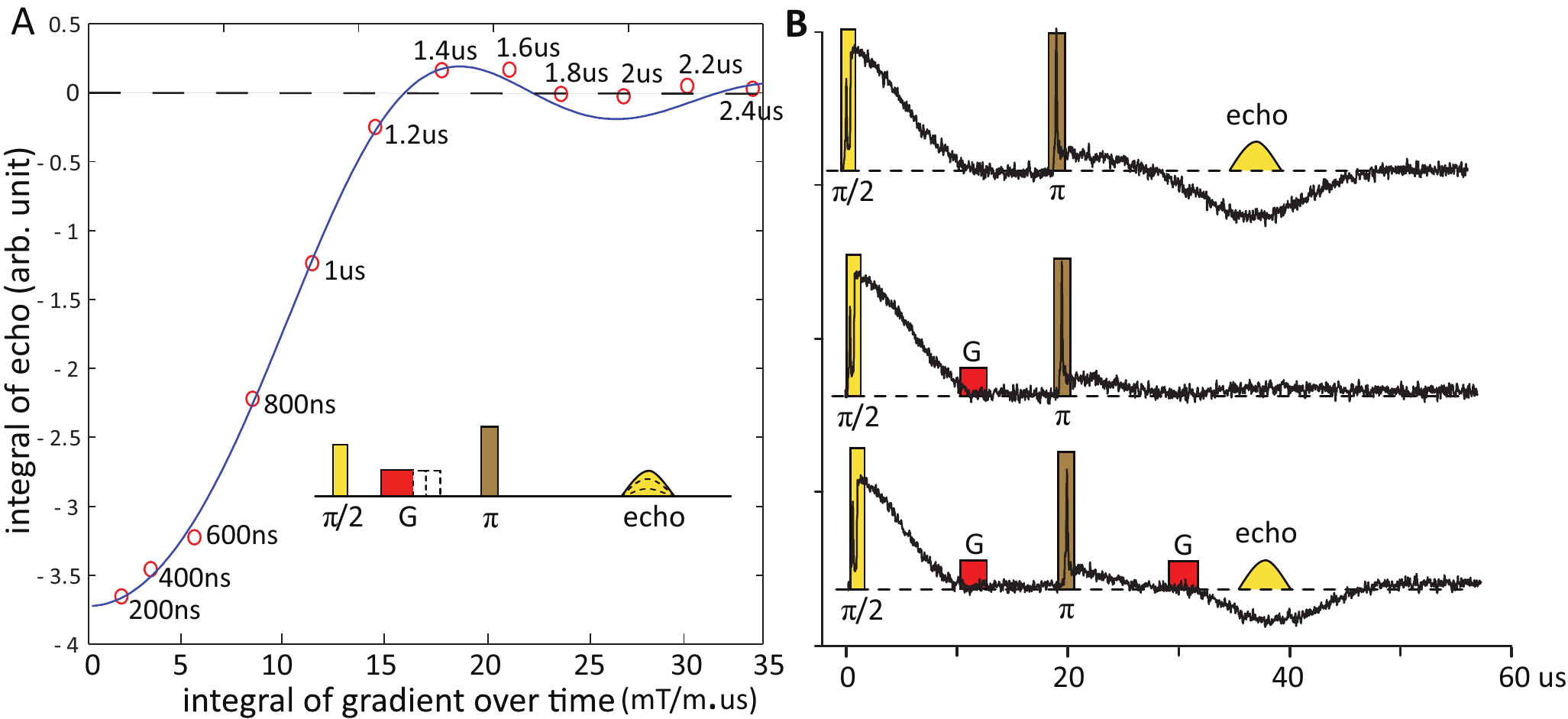}
\caption{\textbf{Single microwave pulse stored in a mode orthogonal to $k=0$ mode across a spin ensemble using a field gradient pulse} (A): The integrated intensity of the spin echo is plotted against the integral of the amplitude of the gradient pulse over time. Considering the profile of the sample, the fitted curve is of the form $B_1(kr_0)/kr_0$ (see Methods), corrected with a second order radial inhomogeneity of the microwave field. Points of zero echo intensity correspond to modes orthogonal to $k=0$ mode. (B): A second identical gradient pulse is applied after the refocussing $\pi$ pulse to shift the stored pulse back iinto $k=0$ mode, which then radiates to give an echo. Sample: $^{14}$N$@$C$_{60}$ at room temperature.}\label{f1}
\end{figure}

Our first experimental objective is to demonstrate that by applying an appropriate magnetic field gradient we can create an orthogonal global phase mode across the sample. By varying the duration $\tau$ (or alternatively the amplitude $G$) of the field gradient pulse, we create modes of varying $k$, resulting in different couplings to the radiative $k=0$ mode. When we insert such a field gradient pulse into a Hahn echo sequence we observe a variation in the spin echo intensity of the N@C$_{60}$ sample, as shown in Fig. \ref{f1}A. A disappearance of the spin echo indicates that the uniform phase coherence of the spins is completely eliminated, and that the mode $k$ generated by the field gradient pulse is orthogonal to mode $k=0$. These results are used to obtain an appropriate duration and magnitude of the gradient pulse (e.g.~$\tau=1.3~\mu$s and $\sim30$ mT/m) that will be used in the subsequent experiments for storing multiple microwave excitations.

The elimination of phase coherence by using magnetic field gradients has been exploited in NMR experiments \cite{Teklemariam2001, Teklemariam2002} where the field gradient was used as a phase coherence eraser. Here, however, we use the pulsed field gradient to encode and label excitations stored in the multi-mode memory. In Fig. \ref{f1}B we show that a single microwave excitation generated by a microwave $\pi/2$ pulse and stored into a $k$-mode orthogonal to the $k=0$ mode, can be retrieved by applying the same field gradient pulse after the refocusing $\pi$-pulse. The recovered spin echo appears in the same place as for a standard Hahn echo sequence, with a fidelity of $92\%$.

%\subsection{Storage of multiple microwave excitations}

%\subsubsection{Storage of multiple excitations using pulsed field gradient}\label{sec2}

A key resource for both classical and quantum computation is a memory allowing reading and writing of registers. The memory is particularly useful if registers are accessible in arbitrary order. In our scheme, we achieve this by converting the register to be accessed to the $k=0$ mode using appropriate gradient pulses. The gradient pulses also shift all other registers in $k$-space by the same amount; the memory can be thought of as a ``Turing tape'' residing in $k$-space. 

\begin{figure}
\includegraphics[width=8cm]{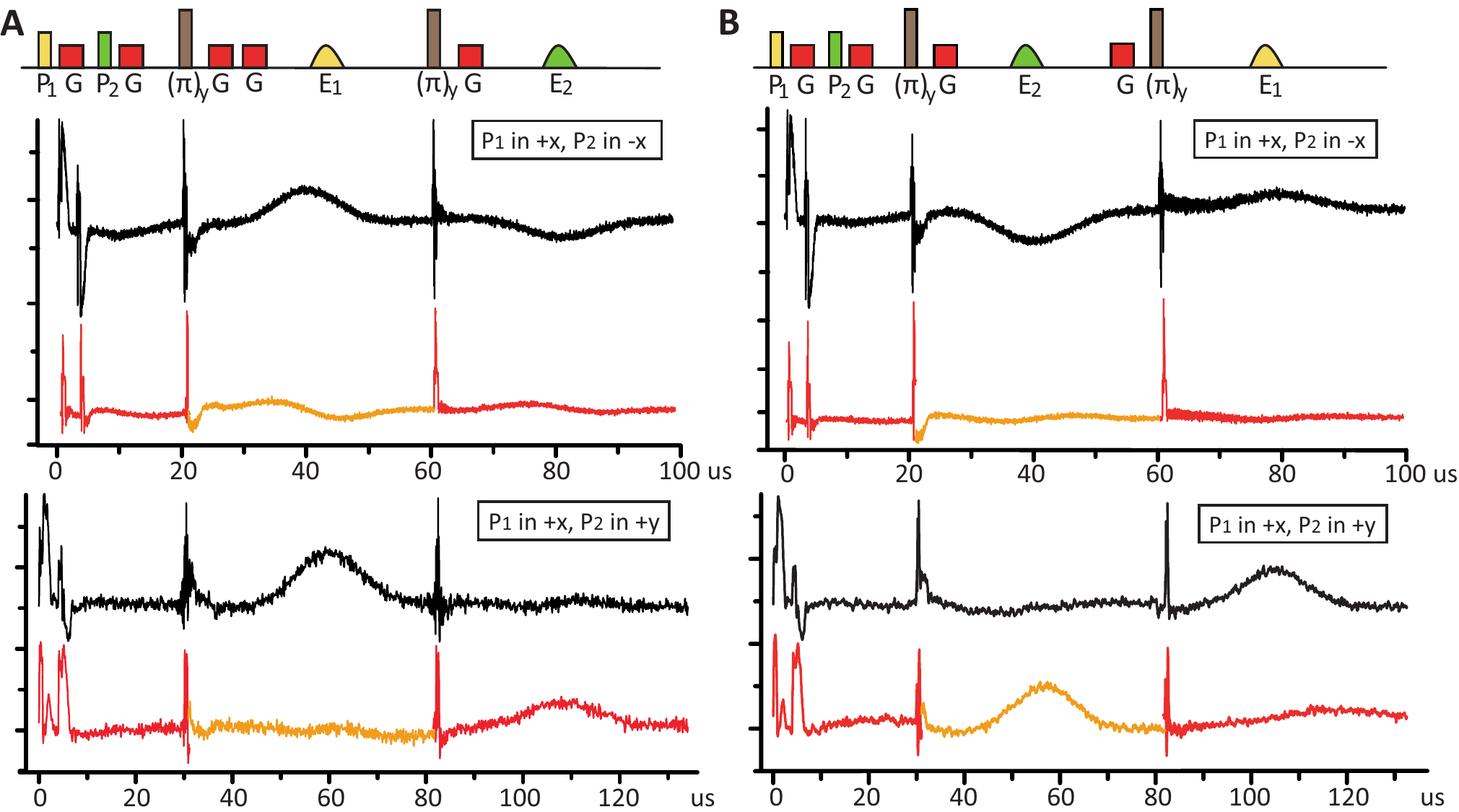}
\caption{\textbf{Recalling two stored microwave pulses in an arbitrary order using pulsed field gradient} (A): Two microwave pulses recalled in the same order as they are stored. (B): Two pulses recalled in the inverse order. Transients are shown respectively for P$_1$ and P$_2$ applied in $+x$ and $-x$, $+x$ and $+y$ direction. In each of the four diagrams we show the real part ($+x$ direction, black) and imaginary part ($+y$ direction, red) of the signal. As the refocusing pulse is in $y$ direction, the imaginary part of the signal between the first and second refocusing pulse (orange) is inverted to keep the echo and free induction decay (FID) having the same sign. In this experiment P$_1$ and P$_2$ are $\pi/6$ pulses. They are kept relatively strong so that we can observe a clear echo. Sample: $^{14}$N$@$C$_{60}$ at room temperature.}\label{f4}
\end{figure}

Figure~\ref{f4} shows an experiment in which we store two weak microwave excitations of different phase (P$_1$ and P$_2$) in the spin ensemble and recall them in either order. Each excitation is followed by an encoding field gradient pulse $G$. Since the field gradient pulses act on all the microwave excitations stored in the ensemble, the first excitation P$_1$ is effectively stored into the mode $2k(G,\tau)$ while the second P$_2$ is stored in mode $k(G,\tau)$. Therefore, after the first refocusing $\pi$-pulse, we can restore and refocus either P$_1$ or P$_2$ to the $k=0$ mode by applying either two or one gradient pulses respectively. This encoding method can be readily generalized to the case of multiple microwave excitations: by using pulsed field gradients we can address any of the stored excitations and retrieve them individually. 

%\subsection{Crosstalk between two stored excitations}

In the limit of single quantum excitations stored in orthogonal $k$-modes, there is no crosstalk between stored registers. However, in our experiment the excitations that we store and retrieve are far from single quantum excitations. Therefore, we investigate the crosstalk between two stored microwave excitations as a function of their intensities. We compare the echo intensity $I$ of one recovered excitation in the presence of another excitation with the echo intensity $I_0$ in the absence of another excitation, and calculated the fractional change $D=(I_0-I)/I_0$. Supposing that the two microwave excitations are stored in two orthogonal modes, we would expect a decrease in the echo of the second excitation (P$_2$) when increasing the intensity of the first excitation (P$_1$) because the net magnetization along the $z$ direction is partly consumed by P$_1$, leading to $D_2=1-\cos(\theta_1)$ where $\theta_1$ is the tipping angle for P$_1$. We would also expect a decrease in the echo of P$_1$ with the increase of the intensity of P$_2$ due to the partial refocusing effect of P$_2$, which is a result of the fact that {\em any} two tipping pulses generate some refocusing (even if the second pulse is not a $\pi$-pulse as required for optimal refocusing). This leads to $D_1=[1-\cos(\theta_2)]/2$, where $\theta_2$ is the tipping angle for P$_2$. We found that $D_1$ and $D_2$ follow the expected dependencies within 0.5\%, for $\theta_1$ and $\theta_2$ each varying independently in the range 0 to $\pi/2$, confirming that we have identified the key crosstalk mechanisms (see Supplementary Information). Since the error $D$ depends on the tip angles to second order for both mechanisms, it quickly becomes negligible for small tip angles.

Apart from this crosstalk, errors in the echoes can also arise from errors in the microwave pulses \cite{Morton2005}. An inaccurate refocusing $\pi$ pulse will introduce extra excitations in the spin ensemble and pollute the stored information, but there are well established methods for mitigating the dominant instrumental imperfections~\cite{Morton2005a}. Alternatively, such errors could be avoided if we were able to invert the inhomogeneity without using microwave pulses. Techniques of this kind have been developed for multi-mode memories with atomic ensembles, e.g.~CRIB and atomic frequency comb~\cite{Afzelius2009}.

%\subsubsection{Storage of multiple excitations by constant field gradient}\label{sec1}

\begin{figure}
\includegraphics[width=8cm]{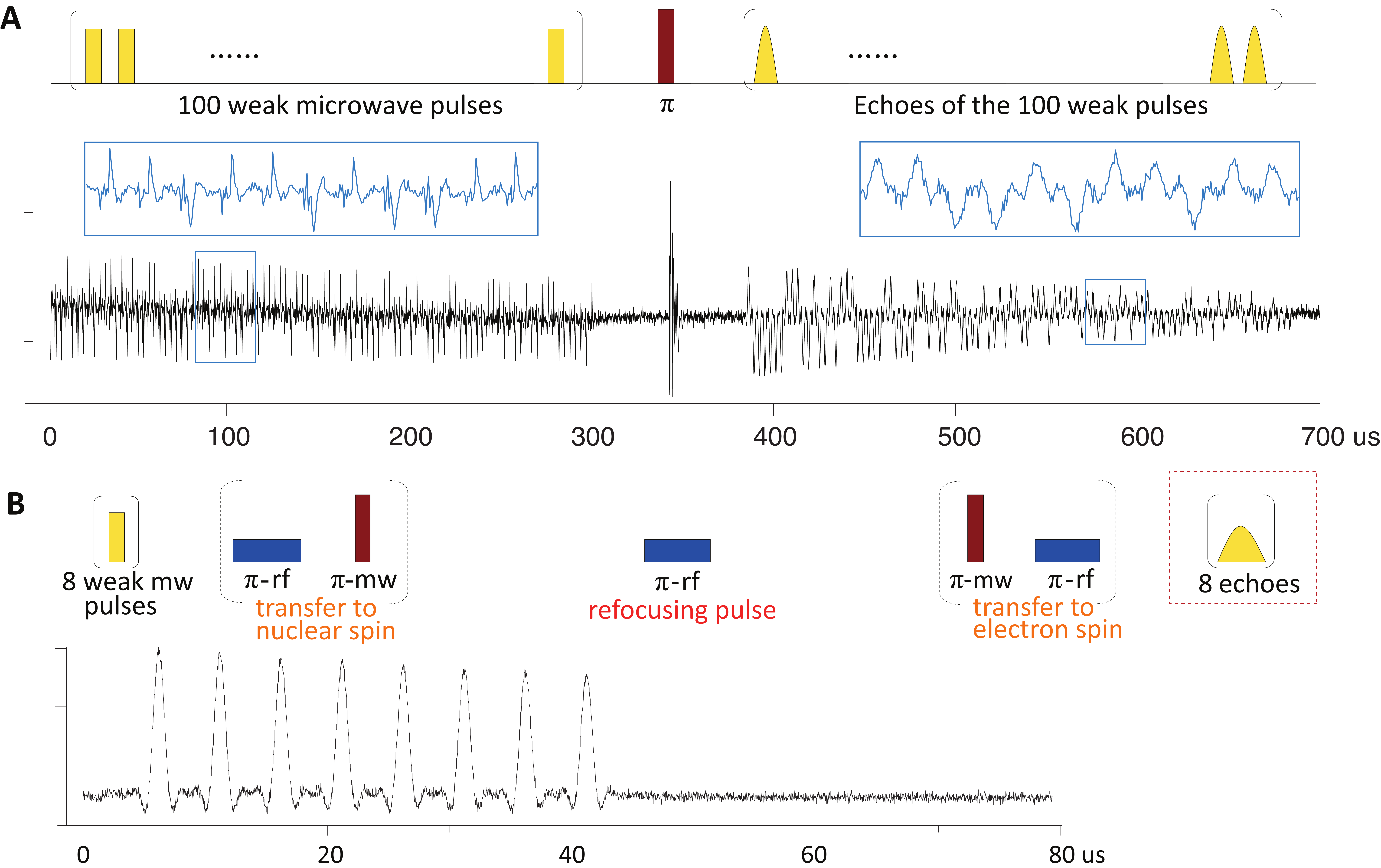}
\caption{\textbf{Multiple microwave pulses stored in multi-modes of the ensemble using constant field gradient} (A): Storage of 100 weak microwave pulses. Each pulse has a flip angle of less than 0.01 $\pi$ and is applied arbitrarily in the $+x$ or $-x$ direction. The FID of the sample is shorter than 2 $\mu$s in this constant field gradient, so the microwave pulses are separated by 3 $\mu$s. After the refocusing pulse the stored pulses are recovered in reverse time order. (B): 8 weak microwave pulses are stored in the electron spin ensemble using constant gradient. All the different $k$ modes are transferred into the nuclear spin, after applying an rf $\pi$-pulse, transferred back to the electron spin. The transient shows the 8 echoes (enclosed in the red dashed box in the pulse sequence) observed corresponding the 8 stored excitations. Sample: P-doped $^{28}$Si at 9~K.}\label{f3}
\end{figure}

Having demonstrated that we are able to store multiple excitations and that the crosstalk can be kept small for weak excitations, we address the question of how many excitations it is practically possible to store. In this experiment we apply a static field gradient to an ensemble of P-donor spins in silicon by placing a small permanent magnet close to the resonator, and apply a train of weak (less than 0.01$\pi$) excitations at intervals in time such that each excitation becomes orthogonal to the $k=0$ mode before the application of the subsequent one, i.e.~the time separation of pulses exceeds the ensemble dephasing time $T_2^*$. The microwave pulses are much shorter than $T_2^*$, ensuring that the bandwidth is sufficient to excite the whole ensemble uniformly. Fig.~\ref{f3}A shows the storage and retrieval of the 100 weak microwave pulses individually. The pulses are applied arbitrarily in $+x$ and $-x$ direction to represent a register of information. As the field gradient is constant, only one refocusing $\pi$-pulse is sufficient to recall all the stored excitations. The wave number register forms a quantum memory \emph{stack}, and the excitations are recalled in reverse order of there storage~\cite{Hahn1955}. 

The memory time of the information is limited by the $T_{2}$ decoherence time of the electron spin (about 450 $\mu$s for P-donor spin at 9K), as seen in the decrease of the intensity of the echoes in Fig.~\ref{f3}A. On the other hand, the shortest separation between each microwave excitation is the $T_2^*$ in the presence of the gradient (in this case 2 $\mu$s). These two figures set the limit for the number of microwave excitations that can be stored in the ensemble. Although a memory time of several hundreds of microseconds is already much longer than the best superconducting qubits, and times as long tens to hundreds of milliseconds have been reported~\cite{Tyryshkin2003,Tyryshkin2006,Tyryshkin2009}, the stored excitations can be transferred to nuclear spin in order to benefit from even longer nuclear spin coherence times in excess of seconds~\cite{Morton2008}. Fig.~\ref{f3}B shows the storage of eight weak microwave pulses using a constant field gradient. We then transfer the entire state of the electron spin ensemble (i.e.~all the microwave excitations stored in different $k$ modes) simultaneously to a nuclear spin degree of freedom. That this transfer is a transition between hyperfine states of each individual atom and it can hence be carried out on a time scale given by the hyperfine interaction strength, which is in the order of microseconds in our system. As the excitations are now in nuclear spins, we achieve the refocusing with a radio frequency $\pi$-pulse resonant with the nuclei instead of a microwave $\pi$-pulse. We then transfer the coherences back to the electron spin and echoes of the eight original microwave excitations are observed.

Our experiments with small tipping angles demonstrate how a single electron spin ensemble could be used as a multi-mode solid-state memory. The prerequisite techniques for extending this work to the quantum regime, synthesizing arbitrary single photon states~\cite{Hofheinz2009} and non-destructively detecting single cavity photons~\cite{Johnson2010}, have now been achieved in experiments with superconducting qubits.  Recent experiments have also demonstrated that large couplings can be achieved between spin ensembles and cavities~\cite{Schuster, Saclay}. Further studies are necessary to show that the demonstrated coherence of the qubits and spins can be maintained in a combined system.

%\section{Conclusion}
In summary, we have demonstrated holographic storage of microwave pulses in an electron spin ensemble. By using magnetic field gradients the microwave pulses are encoded in different collective modes of the spin ensemble, and are retrieved individually. More robust storage can be achieved by transferring the coherence from electron spin to a coupled nuclear spin. These results show the prospect of using spin ensembles as a memory medium and implementing a quantum computing scheme with hybrid physical systems.

HW is supported by KCWong Education Foundation; KM by the EU integrated project SCALA; AA and JJLM by the Royal Society; JHW and GADB by QIPIRC (EPSRC GR/S82176/01, EP/D074398/1); RJS and DIS by NSF DMR-0653377 and Yale University; KMI by MEXT \#18001002, FIRST, Nanoquine, and  JST-EPSRC SIC Programs; and JJLM by St. John's College, Oxford; and JHW by the National Research Foundation \& Ministry of Education, Singapore.

\newpage
%\begin{widetext}
\include{supple-info}
%\end{widetext}

\end{document}

%% file: supple-info.tex
%\documentclass{article}

%\usepackage{amsmath,amsthm,amsfonts,amssymb,bm}
%\usepackage{graphicx}

%\begin{document}

\begin{center}
Supplementary Information
\end{center}

In Fig.~4 we show the crosstalk between different modes $k$ as a function of the intensity of the stored microwave excitations. We plot both experimental data and theoretical calculations. The curves for the theoretical results are $D_1=(1-\textrm{cos }\theta_2)/2, D_2=1-\textrm{cos }\theta_1$, with $\theta_1$ and $\theta_2$ the flip angles of P$_1$ and P$_2$. The deviation between the experimental data and the theoretical curves is less than 0.5$\%$.

\begin{center}
\begin{figure}[h]
\includegraphics[width=8cm]{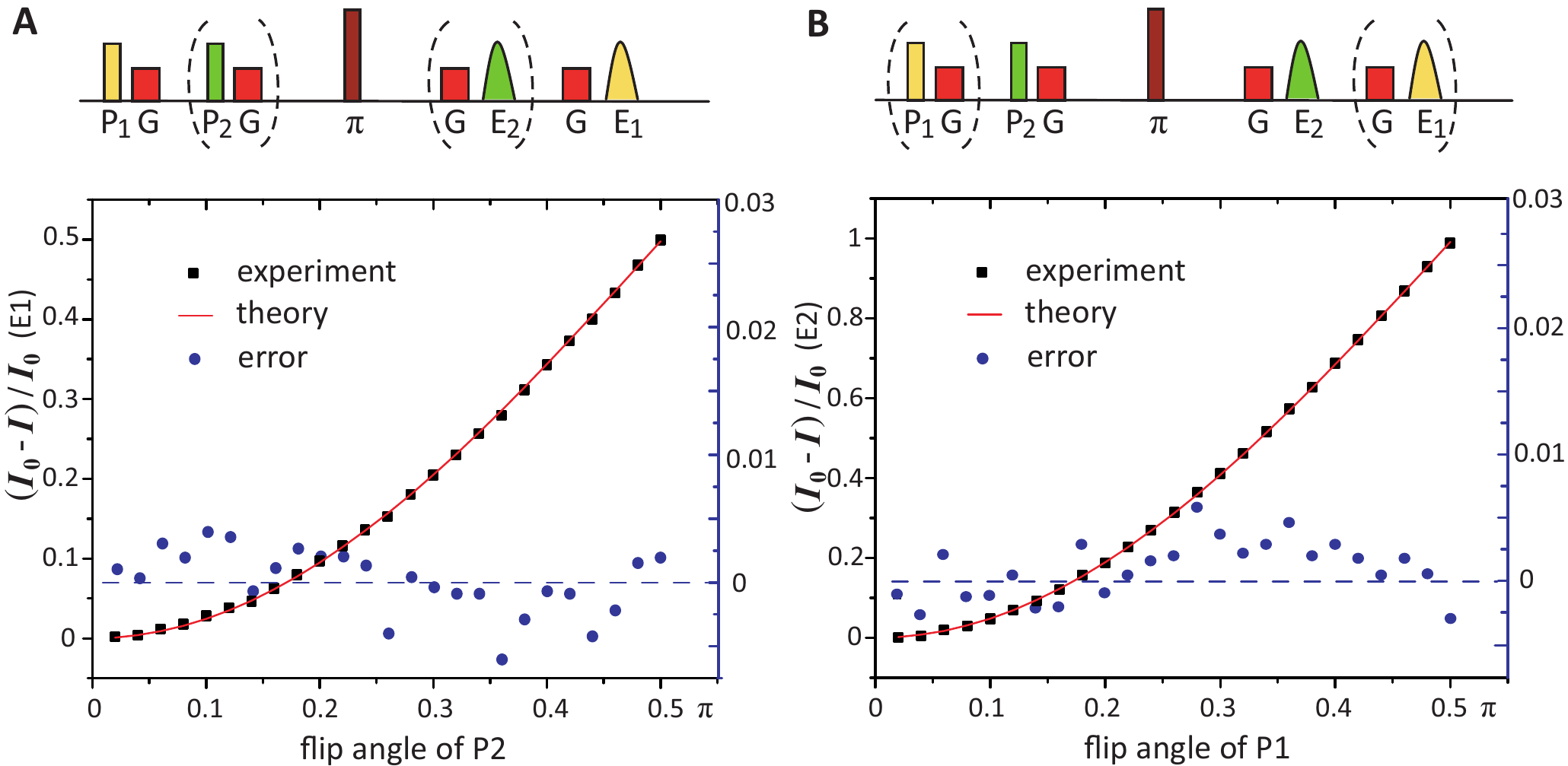}
\caption{\textbf{Difference in the recovered echo intensity ($D$) of one microwave excitation caused by another excitation} (A): $D$ for the echo of the first microwave pulse (P$_1$) E$_1$ as a function of the flip angle of the second microwave pulse P$_2$. (B): $D$ for E$_2$ as a function of the flip angle of P$_1$. In both cases the durations of P$_1$ and P$_2$ are swept from less than $0.02\pi$ to $0.6\pi$, however $D$ was independent of the intensity of the microwave pulse whose echo intensity was measured. }\label{f5}
\end{figure}
\end{center}

%\end{document}